\pdfoutput=1 %
\documentclass[sigplan,nonacm,table,prologue]{acmart}
\renewcommand\footnotetextcopyrightpermission[1]{}
\PassOptionsToPackage{hyphens}{url}
\usepackage[shortcuts]{extdash}  %
\usepackage{multicol} %
\usepackage{tabularx, url} %
\usepackage{graphicx} %
\usepackage{enumitem}
\usepackage{hyperref}
\usepackage[many]{tcolorbox}
\usepackage{makecell}
\usepackage{xspace}
\usepackage{amsmath}
\usepackage{xparse}
\usepackage{array}
\usepackage{dblfloatfix}
\usepackage{listings}
\definecolor{Purple}{HTML}{A020F0}
\usepackage{fancyvrb}
\usepackage{verbatimbox,caption,float,lipsum}
\usepackage{multirow}
\usepackage[labelfont=bf,font=small]{subcaption}

\usepackage{natbib}
\usepackage{amssymb}
\usepackage[prologue,table]{xcolor} %
\usepackage[capitalize, nameinlink]{cleveref}
\crefformat{section}{#2\S{}#1#3}
\Crefname{section}{Section}{Sections}
\makeatletter
\Crefformat{section}{\Cref@section@name~#2#1#3}
\makeatother

\usepackage{tikz}
\newcommand{\cnumber}[1]{\tikz[baseline=(myanchor.base)] \node[circle,fill=.,inner sep=1pt] (myanchor) {\color{-.}\bfseries\footnotesize #1};}

\newcolumntype{P}[1]{>{\centering\arraybackslash}p{#1}}
\newcolumntype{M}[1]{>{\centering\arraybackslash}m{#1}}
\newcolumntype{L}[1]{>{\raggedright\arraybackslash}m{#1}}

\definecolor{darkblue}{RGB}{0, 51, 102}
\definecolor{mygreen}{RGB}{0, 150, 0}
\definecolor{myred}{RGB}{200, 0, 0}

\newcommand{\draft}[1]{}
\begin{document}

\date{}

\title{\Large \bf Extending eBPF observability to Non-standard execution environments.}

\author{Pamenas Kariuki}
\affiliation{%
  \institution{TU Dresden}
  \country{Germany}
}
\author{Andre Martin}
\affiliation{%
  \institution{TU Dresden}
  \country{Germany}
}
\author{Christof Fetzer}
\affiliation{%
  \institution{TU Dresden}
  \country{Germany}
}

\pagestyle{plain}
\begin{abstract}
    eBPF observability of non-standard execution environments (NEEs) like TEEs or LibOSes is hindered by their unconventional exception-handling and memory-access mechanisms that limit standard Linux tooling. This work introduces two mechanisms that enable eBPF-based observability for NEEs: (1) kernel memory extensions for safely accessing NEE memory from eBPF programs, and (2) a lightweight flexible probe performance measurement unit (LWFP PMU) that provides flexible and generic probing, for NEEs, through the following LWFP: simple (SLWFP), enclave (ELWFP) and extended (ExLWFP) probes.

We demonstrate the practicality of these extensions by developing tooling for tracing, stack sampling with Flame Graphs, dynamic instrumentation, timing analysis, and USDT support for Intel SGX enclaves and LibOSes.
Performance measurements show that SLWFP probes achieve a latency of 394\,ns, outperforming
uprobes, which exhibit 25\% higher latency, while ELWFP probes incur a latency $\sim$2.8\,
\textmu s, which is practical for enclave observability.
The addition of SLWFP introduces negligible overhead to existing uprobe performance.
Taken together, this work lays the foundation for closing the long-standing gap between 
NEE tooling and Linux observability tooling by enabling generic and reusable eBPF tooling
for diverse NEE hardware and software architectures.

\end{abstract}
\maketitle

\thispagestyle{empty}

\section{Introduction}

The development and shift towards cloud computing motivated the rise of OS- and VM-based containerization technologies to address the need for lightweight isolation, efficient resource sharing, and scalable multi-tenant deployments. However, despite their success, these approaches left important gaps in flexibility, trust minimization, and interface specialization. Library operating systems (LibOS) architecture, including unikernels, rose to fill this gap: small Trusted Computing Base~(TCB), broader application compatibility and personalization, non-native interface support, low attack surfaces, and stronger multi-tenant isolation~\cite{engler1995exokernel,porter2011drawbridge,porter2014graphene,hunt2016ryoan,bittau2009nexus}.

Cloud security was addressed by trusted execution environments (TEEs), that provide confidentiality and integrity guarantees to software. Popular TEEs include AMD SEV-SNP~\cite{kaplan_amd_nodate}, Intel TDX~\cite{intel2022tdx}, and Intel SGX (SGX)~\cite{costan_intel_2016}. SGX has a radical programming model that splits an application into an untrusted part and a trusted part in an enclave. An enclave is protected from observation by the host OS. LibOSes were adopted as a practical solution to SGX programming challenges.

SGX, LibOS and unikernels rely on non-standard mechanisms—e.g., loading and executing code, exposing CPU state, and handling exceptions—that are opaque to the host OS. This situation extends to SGX-like TEEs built on similar principles as well as VM unikernels. In general, there are compute extensions and hardware that allow isolated and protected memory ranges to be mapped into a process memory, and entail the aforementioned features. We collectively categorize such software and hardware execution environments as Non-standard Execution Environments (NEEs).

\subsection*{The Observability Problem}
Optimizing performance, debugging, testing and achieving correctness for software running with restricted interfaces presents a significant challenge. This fundamentally comes from different basic assumptions present NEEs in contrast to native environments.
Further, the problem is aggravated by tooling gaps: tools and infrastructure to identify issues, gather evidence and affirm correctness are relatively lacking in comparison to native tools.
Although a part of this problem can be countered by hardware~\cite{mckeen_intel_nodate,intel_xeon_6979p_specs} and software architectural enhancements ~\cite{al_scone_2016,noauthor_gramineprojectgramine_nodate,baumann_shielding_2015,shen_occlum_2020}, it is still insufficient to address all aspects of observability, performance and correctness.

Introspection is commonly employed to fill the tooling gap leading to repetition among the individual frameworks. In some restricted environments--such as production mode SGX--this is the only recourse. However, in debug and pre-production environments, as well as in software NEEs, there is greater freedom to employ other techniques for richer observability of NEEs.
 
Practitioners deploying NEEs need to answer critical observability questions to characterize and optimize performance:
\begin{itemize}
  \item \textit{Where is execution time spent?} Which functions consume the most CPU cycles?
  \item \textit{What are the bottlenecks?} Is performance limited by transition latency, syscall overhead, memory access patterns, or application logic?
  \item \textit{How does actual execution diverge from expected behavior?} Are unexpected functions being called? Is library behavior inefficient?
  \item \textit{What evidence can we gather to test bug and performance hypotheses?}
  \item \textit{Can we prove software correctness persists under different environment?}
\end{itemize}

These questions require observability tooling capable of inspection of NEE execution, especially those capable of operation with minimal or ideally without modifying applications or recompiling of software.

\subsection*{Why Existing Tools Fail}

Several existing observability tools target NEEs, but each has fundamental limitations. Intel VTune~\cite{intel_vtune_2026} and the GDB plugins~\cite{noauthor_sgx_gdbplugin} require software recompilation and introspection or instrumentation libraries embedded in code.
 Most LibOS and unikernels solve this problem by introducing introspection libraries. However, this approach requires implementation for each library, duplicating efforts; a strong contrast with reusability of native tools.
On the other hand, there are narrow tooling implementations: Research tools like TEE-Perf~\cite{bailleu_tee-perf_2019} and SGXoMeter~\cite{mahhouk_sgxometer_2021} target specific metrics (latency breakdown, performance counters) and cannot provide general-purpose tracing or dynamic instrumentation.

Native Linux has a rich set of observability tools—perf, strace, uprobes, and kprobes— but fail to observe NEEs on account of the following limitations:

\begin{itemize}
  \item Probing limitation. \textit{user probes} rely on the \texttt{mmap} syscall to detect code loading. When the kernel observes \texttt{mmap}, it scans virtual memory areas and installs INT3 breakpoints at probe points. However, NEE code is loaded via direct memory copying of code into memory, without triggering \texttt{mmap}. Consequently, uprobes never install.
  
  \item \textit{eBPF programs} cannot access protected memory in hardware NEEs as special instructions are needed for memory access. eBPF helpers fail silently making eBPF-based profiling and tracing are unavailable for NEEs.
  
  \item \textit{Hardware breakpoints} are fixed per CPU core (4 in X86), limiting the number of simultaneous addresses that can be probed, although their programming is more flexible that software breakpoint with respect to NEEs.
\end{itemize}

\subsection*{Our Solution: Three Primitives for NEE Observability}

In order to address these limitations, we introduce the following three primitives:

\noindent\textbf{P1: Safe Memory Access from eBPF.} We extend eBPF memory helpers to recognize and handle special memory regions via new VMA (Virtual Memory Area) operations. 
A hardware NEE driver implements the new methods to support this feature. This enables eBPF programs to access and modify NEE state directly.

\noindent\textbf{P2: Lightweight Flexible Probes (LWFP).} 
 We introduce, a new LWFP Performance Monitoring Unit (LWFP PMU) associated with LWFP.
These software-based probes relieve the probe discovery and insertion roles from the kernel, leaving only exception reporting and probe management.
Further, they introduce a context-aware matching of execution context to enable special needs for NEEs like SGX.

\noindent\textbf{P3: NEE-Aware Exception Handling.} We integrate LWFP with Linux's exception and eBPF infrastructure, collectively allowing 

Together, these three primitives enable eBPF observability tools to work across NEEs with the same ease and low overhead as on native Linux, without requiring application recompilation or embedded instrumentation libraries.

\subsection*{Contributions}

This work makes the following contributions:

\noindent\cnumber{C1} \textbf{Problem analysis}: We present a systematic analysis of limitations of general observability of NEEs in Linux, identify the core architectural constraints and their respective manifestations across different NEE types.

\noindent\cnumber{C2} \textbf{eBPF memory extensions}: We design and implement kernel extensions that enable eBPF programs to safely access protected memory regions via pluggable backend drivers and demonstrating the capabilities with SGX.

\noindent\cnumber{C3} \textbf{Lightweight Flexible Probes}: We develop a new perf PMU that supports dynamic probe insertion and context-aware matching across NEEs, enabling zero-overhead probe placement independent of memory layout or initialization method.

\noindent\cnumber{C4} \textbf{Performance evaluation}: We demonstrate that LWFP achieve latency of 397 ns for native LWFP probes (26\% faster than uprobes, 17.5\% faster than hardware breakpoints) and ~2.8 $mu$s for enclave-targeted probes, establishing practical overhead.

\noindent\cnumber{C5} \textbf{Observability tooling}: We build eBPF-based tools for CPU profiling (DWARF stack sampling), system call tracing, function timing, and USDT instrumentation, demonstrating the generality, practicality and utility of the primitives across multiple observability use cases on real-world applications.

\section{Background}\label{section:background}
In this section, we provide relevant technical information to elucidate the observability problem of general NEEs and SGX  in Linux.

\subsection{Intel SGX }\label{subsection:intel-sgx}
Intel Software Guard Extensions (SGX)\cite{costan_intel_2016} creates protected memory regions called enclaves, that are isolated from the OS. Each enclave is associated with several data structures, with the most relevant ones  being presented hereafter.

The \textbf{SGX Enclave Control Structure (SECS)} defines enclave metadata, 
including its base address, size, State Save Area (SSA) frame sizes,  and global attributes. Multithreading is managed via a per-thread structure, Thread Control 
Structure (TCS), which specifies entry and  asynchronous enclave exit (AEX) addresses, settings,  and active SSA frames.

\textbf{SGX exception flow}.\label{SGX:exception-flow}
\textit{AEX} occurs when interrupts or exceptions occur while executing enclave. The CPU saves the extended state into the next available \texttt{SSA} frame and replaces it with an AEX-defined synthetic state. Specifically, \texttt{RAX} is loaded with the \texttt{ERESUME} leaf value, \texttt{RBX} stores the current \texttt{TCS} address, and both \texttt{RIP} and \texttt{RCX} point to the Asynchronous Exit Pointer (AEP). AEP is a trampoline code block containing an \texttt{ERESUME} instruction to reenter the enclave post-handling.

\begin{figure}[hbtp!]
    \centering\includegraphics[width=0.4\textwidth]{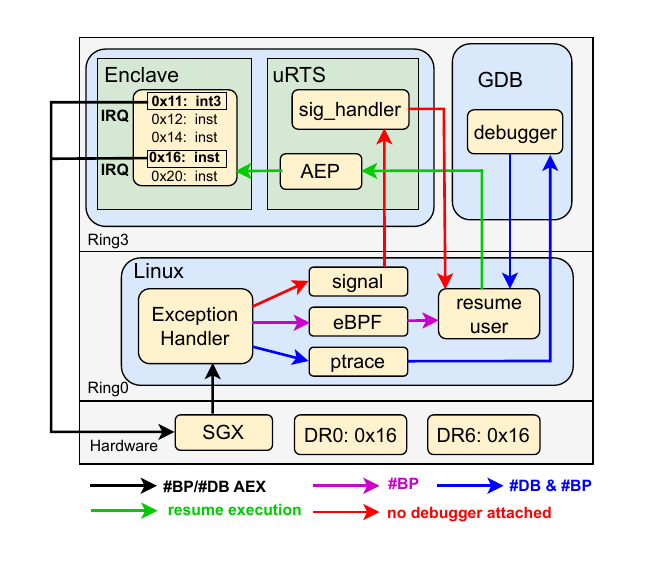}
    \caption{SGX AEX on debug and Linux exception reporting}
    \label{fig:linux-sgx-mechanism}
\end{figure}

\textbf{Enclave behavior on debug}.
Enclave debug is enabled in the \texttt{SECS} and for each \texttt{TCS} which permits permits single-stepping, memory inspection via \texttt{EDBGRD} and \texttt{EDBGRW}, breakpoint and events reporting to OS. A hardware breakpoint (\textit{HB}) is a fault (\verb|#DB| exception), whereas a software breakpoint (\textit{SB}) is a trap (\verb|#BP| exception). On debug opt-out , both exceptions trigger an invalid opcode exception (\verb|#UD|). \Cref{fig:linux-sgx-mechanism} illustrates this process within the Linux kernel. If no debugger is attached, a signal is delivered directly to the untrusted host application; otherwise, the \textit{ptrace} subsystem forwards the breakpoint events to the debugger. Additionally, \textit{Perf} can leverage an \textit{HB} to trigger the execution of an eBPF program.

\subsection{eBPF}\label{subsection:ebpf}
eBPF is an in-kernel 64-bit RISC virtual machine that executes sandboxed user code to extend kernel functionality without custom modules or kernel modifications. eBPF is integrated with the \texttt{bpf} system call and event-driven hooks, allowing its application for packet-filtering use cases, security~\cite{noauthor_filter_nodate_seccomp}, and system observability~\cite{noauthor_parca-devparca_2024}.

To ensure safety, a kernel verifier enforces program termination via limited recursion and loop unrolling. To bypass its strict 1-million-instruction restriction limit, eBPF supports BPF-to-BPF function calls and tail calls to alternative programs . Additionally, the \textit{bpf\_loop} helper facilitates large iterations of up to 1 million loops.

\textbf{eBPF programs and data structures.}
The \texttt{bpf} system call manages program loading and the lifecycle of eBPF maps (e.g., hashmaps, arrays, and program arrays). These maps are shared across the user-kernel boundary for efficient, zero-copy data transfer and storage~\cite{cassagnes_rise_2020}. Loaded programs are referenced via file descriptors (\textit{fd}) and linked to kernel event sources using \textit{ioctl} for \textit{perf\_event} subsystems.

\textbf{eBPF helpers.}
Kernel-defined helper functions expose stable APIs for memory access and complex operations. They guarantee memory safety by encapsulating low-level routines; for instance, \textit{bpf\_probe\_read} wraps around the kernel's \textit{copy\_from\_user\_nofault} routines, which use \textit{\_\_put\_user\_size} for direct RAM access. Consequently, eBPF programs cannot issue specialized I/O instructions or access memory managed exclusively by third-party device drivers.

\textbf{eBPF userspace ecosystem.}
Modern infrastructure simplifies eBPF development and runtime management~\cite{noauthor_ebpf_nodate}. This work leverages frameworks like BPF Compiler Collection (BCC)~\cite{noauthor_iovisorbcc_2025} for high-level language interface alongside an LLVM-backed C toolchain, as well as \textit{libbpf}~\cite{noauthor_libbpflibbpf_2026} for native C and Rust program orchestration.

\subsection{Linux Observability Infrastructure}\label{subsection:linux-debug-pmu-infrastructure}
This section presents the main tools and interfaces for observability and tooling in Linux OS.

\subsubsection{Ptrace}\label{subsection:ptrace_background} ~\cite{noauthor_ptrace2_nodate,keniston_ptrace_nodate} is the core system call for controlling execution of inferior processes for tracing  and debugging. Ptrace requires several context switches between kernel, inferior, and tracer making it characteristically slow. 
Common ptrace-based tools include: \emph{strace}~\cite{noauthor_strace1_nodate} for tracing issued system calls; \emph{ltrace}~\cite{noauthor_ltrace1_nodate}  tracing library loading; and \emph{GDB}~\cite{noauthor_gdb_nodate} for debugging.

\subsubsection{Linux Perf Events.}\label{subsection:linux_perf_events}
Linux has a performance measurement subsystem that exposes both software and hardware counters through the \emph{perf\_event\_open} system call (PEO). PEO creates a file descriptor~(fd) for a specific performance measurement unit~(PMU), either sampling or counting, that can be read for statistics or as a trigger for execution of eBPF programs. Perf tool is a userspace tool that provides utilities to exploit this interface.

\subsubsection{Linux Probes and Exception Handling}\label{subsection:linux_probes}
Linux provides \emph{kprobes}/\emph{kretprobes} for kernel instrumentation and \emph{uprobes}/\emph{uretprobes} for userspace probing~\cite{mavinakayanahalli_probing_nodate,keniston_kernel_nodate}, managed via the \texttt{bpf} system call. Entry probes execute at function entry, while return probes fire at function exit. For mapped targets, the kernel locates the corresponding Virtual Memory Area (VMA), caches the original instruction, and overrides it with a breakpoint (e.g., \texttt{INT3} on x86). For unmapped binaries, \texttt{uprobe\_mmap} defers activation until a future \texttt{mmap} system call matches the target filename. Because resolution relies strictly on filename matching within the \texttt{mmap} subsystem, executable regions generated dynamically or loaded directly via drivers bypass this mechanism entirely.

When execution hits a breakpoint, the trap triggers the kernel's notifier framework via \texttt{register\_die\_notifier}, passing the exception context to attached handlers like eBPF or \textit{perf} events. To resume the application path post-handling, the cached instruction is executed using either: (1)~\textit{Single-Step In-Line} (SSIL), where the original instruction temporarily replaces the breakpoint under a hardware single-step flag; or (2)~\textit{Single-Step Out-of-Line} (SSOL), where the instruction copy executes within a dedicated memory slot via emulation or direct patching before updating the CPU context.

\section{Design}\label{section:design}
This work is designed to allow flexible instrumentation of code,
eBPF memory access of NEEs and easy integration with other eBPF functionality and tooling.
In  general, in order observable NEEs, the following three primitives are necessary.

\noindent {\cnumber{P1} A means to access NEE memory from eBPF.}

\noindent {\cnumber{P2} A means to flexibly probe (instrument) NEE code.}

\noindent {\cnumber{P3} A means to run a specific eBPF program on firing of an associated target probe.}

\begin{figure}[hbtp]
    \centering\includegraphics[width=\linewidth]{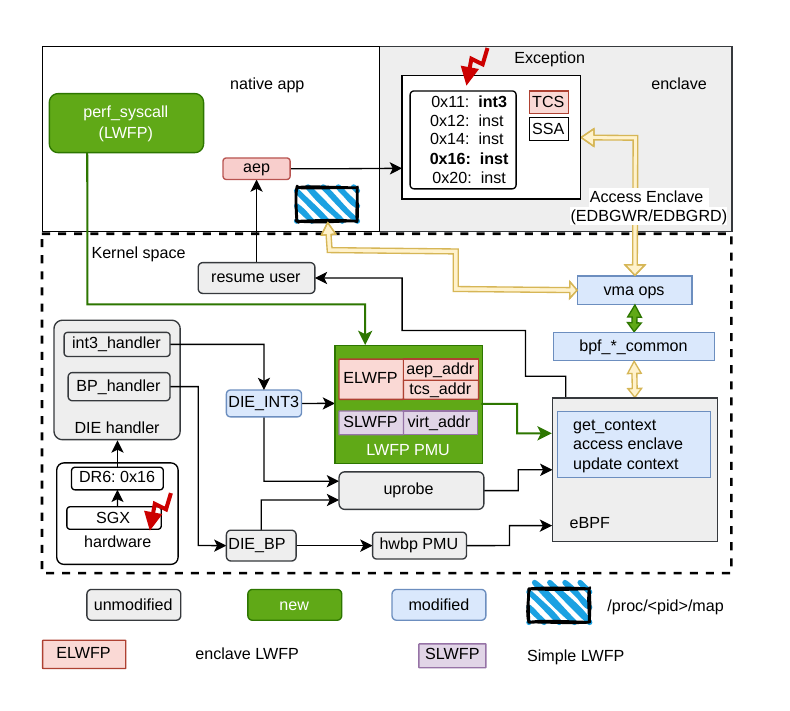}
    \caption{Design overview. New components are shown in green and modified in blue. A new software breakpoint PMU and its probes, ESB and SSB.}
    \label{fig:general-design-overview}
\end{figure}

The following discourse explores the changes made to the Linux kernel to effect
the changes. \cref{fig:general-design-overview} shows an overview of changes made.

\subsection{Extending eBPF Access to NEE Memory}\label{subsection:accessing-enclaves-from-ebpf}
In general, the memory of software NEEs is visible and accessible to bpf programs.
For hardware NEEs, the vanilla Linux kernel requires device drivers to manage and access associated resources.
A common approach and usage of NEEs entails mapping its memory in the address space of a given process while restricting its access with hardware. 
The memory mapping is associated with a virtual memory area.

The goal of memory extension is to allow safe access from bpf programs while maintaining the same exclusive and termination expectations of bpf programs.
This is achieved through two steps: (1) Linux kernel device driver file operations
are extended to provide methods to guarantee safe access or fail fast without triggering faults. And (2) the bpf memory functions are extended to access these
extensions for special VMAs instead of accessing RAM directly.

The driver methods added to  Linux \textit{struct vm\_operations\_struct} allow to do the following: (1) Detect if the VMA backed is a special backend. (2) An access method for the VMA. And finally, (3) methods to request a safe transaction; one that locks and another that unlocks. All methods are required to be transactional.

The bpf memory access functions are extended to check if the VMA is special and if so, safely access the memory. If these methods fail, the bpf functions fall back to original execution paths.
\draft{
    \begin{lstlisting}[language=C, basicstyle=\small]
     const struct vm_operations_struct
     sgx_vm_ops = {
        ...
     	.access = sgx_vma_access,
    +	.lock_safe = lock_safe,
    +	.unlock_safe = unlock_safe,
    +	.operation_allowed = operation_allowed
     };
    \end{lstlisting}
}

\subsection{Lightweight Flexible Probes (LWFP)}\label{subsection:enclave-probes}
LWFP introduce a new architecture and approach to probing. Generally, all the tasks
in the lifecycle management of user-probes are managed by the kernel:
 (1) In-kernel probe registration. (2) Installing the probe in  the target address at an existing file maping or postponing it till the executable is mapped. (3) Delivering notifications and triggering eBPF programs on firing of a probe. (4) Updating architectural registers, and continuing execution after probe handling. (5) De-registration of probes and removal of probes in target memory. 
This model breaks when any fundamental step in the chain is hindered.

LWFP addresses this limitation and, in addition, the fixed number of hardware breakpoints, by splitting the probe management roles in the chain between the kernel and user-space.
The kernel roles in LWFP are as follows: (1) In-kernel probe registration and management. (2) Exception management, delivering of notifications to bpf programs, and aggregation of probe statistics. (3) De-registration of probes.With these scheme, the kernel's role is reduced to probe management and exception delivery instead of entire work:When \#BP events are generated, userspace context is compared to probe registered 
information and matches are then appropriately handled by a new perf PMU, including triggering eBPF or incrementing a counter.

The user-space roles in LWFP are as follows: (1) Probe target location discovery in an NEE (2) probe registration with kernel. (3) Insertion of software debug instruction (INT3 in x86) into a probe address in NEE memory. (4) Advancement of execution context. (5) Probe removal from NEE memory. (6) Deregistration with the kernel.

The split-role design makes LWFP comparatively lighter to \texttt{kprobes} or \texttt{uprobes} and flexible enough to meet the unique features of NEEs.
LWFP is realized through three kernel extensions: (1) The new LWFP performance measurement unit (PMU)
based on software breakpoints presented in \cref{subsubsection:new-software-breakpoint-pmu}. (2) Extensions to kernel exceptions, and bpf handling mechanism illustrated in \cref{subsubsection:exception_handling_and_reporting}. And (3) an extension to the \textit{perf event open syscall} to support LWFP probes discussed in \cref{subsubsection:extending-perf-event-syscall}.

\subsubsection{Extending perf event open syscall(PEO)}\label{subsubsection:extending-perf-event-syscall}

A new PMU identifier, PERF\_LWFP, is introduced to select the LWFP PMU.
allow creation, configuration and connection of probes with bpf programs.
A new PMU identifier is introduced to identify LWFP. 
The design allows creation of three types of LWFP probes:
A simple LWFP~(SLWFP), enclave LWFP~(ELWFP), and extended LWFP~(ExLWFP).

SLWFPs act like traditional probes in that they match the configured value to the instruction pointer at exception.
An ELWFP is targeted towards SGX enclaves and takes two values: an address for the instruction pointer and a value that is matched to an RBX register at exception as described in~\Cref{subsection:intel-sgx}.

ExLWFP are the most general probes that are designed to  match any number of general purpose registers, in addition to the instruction pointer, at exception.
This is achieved by specifying the values of target registers and passing a flag that selects the registered to be matched.  This generality allows them to target any NEE backed that would provide special exception context, like context at VM exit, and collectively describe it as a probe.

For SLWFP and ELWFP, the parameters to configure the probes follow the same patterns as in other probes in \cite{noauthor_perf_event_open2_nodate}. ExLWFP needs an extra structure, passed as a configuration struct, to define the exception context. Together, the aforementioned create LWFP probes for PMU described in~\cref{subsubsection:new-software-breakpoint-pmu}.

\subsubsection{LWFP PMU}\label{subsubsection:new-software-breakpoint-pmu}
LWFP is designed as a counting PMU; a single statistic is generated whenever a probe is triggered.
The core functionality of this module is to manage all registered probes for each process,
search and match registered probes to CPU context when an exception is delivered,
and submit the events to the perf core for further statistics aggregation and/or invocation of eBPF handling.

\textbf{Probes and Events Management.} Management of probes and associated \textit{perf} events is core to the LWFP module. 
A \textit{perf}-event identifier is associated with all Linux kernel \textit{perf} events and, likewise, with a probe. 
The \textit{perf} events are managed by the core \textit{perf} module, while LWFP probes for each event need to be tracked by the LWFP PMU for probe matching and execution. 
This requires a dual approach: managing both \textit{perf} event identifiers and LWFP probes for fast lookups.

In general, LWFP probes are designed to target a particular thread. 
To achieve fast and predictable probe lookups, we utilize hash maps at process and thread granularity, each keyed by the \textit{process id} (PID) and \textit{thread id} (TID) respectively. 
Within a thread, the probes are stored in a hash list for more efficient search during an exception.

To manage fast lookups for LWFP probe reconfiguration, statistics, and removal originating from the \textit{perf} core, we use a global hash map to store LWFP probe addresses with the \textit{perf} event identifiers as keys. 
The combination of both approaches allows low-latency lookups from exception contexts and \textit{perf} core operations.

\subsubsection{Exception Handling and Reporting}\label{subsubsection:exception_handling_and_reporting} 
To aggregate statistics, trigger handlers, or send signals, we need to identify probes at exception and report to the \textit{perf} backend. 
LWFP accomplishes this by registering a callback handler with the Linux kernel mechanism for INT3 notifications.
On exception, this handler is triggered. It matches the TID and PID of the interrupted thread to look for possible probes.
At the thread level, each type of probe is matched against the exception context according to its respective definition.
On success, the probe is submitted to the perf core for statistics aggregation and triggering of  bpf programs by the eBPF backend.

Depending on the NEE, the handling of the exception and advancing execution is the responsibility of the bpf program or userspace following the design outlined in \Cref{subsection:enclave-probes}.
This is particularly the case with ELWFP with SGX enclaves, where the bpf program has to advance the RIP or manage enclave context with extensions in ~\Cref{subsection:accessing-enclaves-from-ebpf}. 
For SLWFP and ExLWFP, context management is done differently.
After executing handlers, the \textit{perf} core returns execution to the LWFP exception handler.

ExLWFP allows context management at this point by using control flags configured for the probe. For instance,
the RIP can be incremented or single-stepped if the respective flags are set. For SLWFP, the kernel normally advances the instruction pointer and handles other flags.
After handling, control is returned to the kernel exception handler for return to normal execution.

\subsection{Design of LWFP BPF Programs}
In this subsection, we discuss the general design of bpf programs for ELWFP probes.
This is illustrative for context management of ELWFP probes and any other NEE that might utilize ExLWFP probes and need special handling at exception.
The core goal is to determine the execution context in NEE hardware and differentiate it from any other closely related context.
For instance, we can differentiate between hardware breakpoints and ELWFP software breakpoints used in SGX and handle each appropriately.
The design considerations for the bpf programs are outlined hereafter.

\textbf{Context discovery}.
SGX applications have to determine if the target thread was interrupted inside or outside the enclave using RIP.
For in-enclave interruption, the correct execution context is found in the SSA, whose address
can be obtained from enclave SECS and TCS address in register RBX as outlined in \cref{subsection:intel-sgx}.
SECS address should be supplied by a userspace frontend or discovered using probes at enclave launch.
Execution context at exception is accessed from SSA using \textit{bpf\_{read/write}} function. This approach can be used by another NEE for the same goal.

\textbf{Handling LWFP probes vs Hardware Breakpoints}.
The programming of hardware registers and exception handling of all HB are handled by the kernel.
For ELWFP, this task has to be handled by the bpf program or a userspace frontend.
 Handling may entail redirection to a trampoline code, for complex probing applications like~\cite{yasukata_zpoline_nodate}, or the RIP in the SSA needs to be incremented or modified as appropriate. 
 These responsibilities are not part of the core of LWFP PMU functionality to 
 allow general and flexible targeting of NEEs.
\subsection{Shared Tooling Mechanisms}\label{subsection:commont-tooling-mechanisms}
This subsection presents shared characteristic aspects of new eBPF tooling using the extensions described previously.
These patterns are encountered when developing bpf programs for NEE using the new extensions.
In addition, these methods are part of the design of the tools that are later presented in section \cref{subsection:tooling}. Some of these aspects can be seen in~\cref{fig:general_structure}.

\subsubsection{Controlling  Target Process}\label{subsubsection:controlling-target}
A tracee can be controlled in one or a combination of the methods below.

\noindent(1) \textbf{kprobes} and \textbf{uprobes} in target executable linked to eBPF programs to 
initialize eBPF maps by hooking into entry or exit of appropriate functions.
This method anticipates getting information during \textit{exec} syscall with kprobes or
when the uprobes in the target binary fire at start up.
This approach can be used to target any, or several instances of the same, application.

\noindent(2) \textbf{ptrace} syscall can launch or attach to an application and obtain its information to initialize eBPF context. It is useful when a launch command is provided or when a target application is already running.

\textbf{Inferior information.}
Relevant data like PID, enclave base address or linkmap can be obtained using (1) and (2)
if the target application or libraries have defined execution or memory addresses that
can be hooked at, and read from. For instance, an eBPF \texttt{strace} program can hook into 
enclave creation functions for Gramine or SCONE to obtain enclave information.
This information can also be supplied by the userspace frontend as described in \cref{subsubsection:frontends}.

\subsubsection{Context Management}\label{subsubsection:context-management}
NEEs running natively will have the correct register state at eBPF program entry.
Other NEEs, like SGX have to obtain it from saved context in architectural defined regions.
SGX establishes its context with \textit{bpf\_read} from SSA  as described in \cref{subsection:intel-sgx} using TCS and SSA addresses as initialized as in \cref{subsubsection:controlling-target}.
For perf timer interrupts, the RBX and RIP distinguishes between in-enclave and native execution context, as in \cref{subsection:intel-sgx}, and determines what context the eBPF program should use.

\subsubsection{Single Stepping}\label{subsubsection:single-stepping}
Probe placement involves replacement of an instruction, at a given VA, with \texttt{INT3}. 
single-stepping-out-of-line (SSOL) is done by having a dedicated eBPF function for each instruction type to be emulated.
The emulation function takes current CPU state, computes and updates affected memory and registers, including advancing RIP appropriately.
The updated state is written back to SSA, in the case of SGX.
Instructions and their arguments are defined in a struct and placed in eBPF hashmaps indexed by
RIP of probes. Userspace frontends are responsible for parsing target instructions
and creating these maps.

\subsubsection{Userspace Frontends}\label{subsubsection:frontends}
Userspace tracers are responsible for acquisition and setup of configuration data in maps and post processing for target application.
eBPF programs should be as small and fast as possible to minimize workload interference.
Tracer application gathers and initializes shared eBPF maps with apt data for the program to work.
Control flow can be altered through shared variables.
Data is collected over shared maps and ringbuffers and all processing left to userspace
where myriad of libraries and tools exist for this task.

\subsection{Tooling}\label{subsection:tooling}
\begin{figure}[hbtp]
    \centering
    \includegraphics[width=\linewidth]{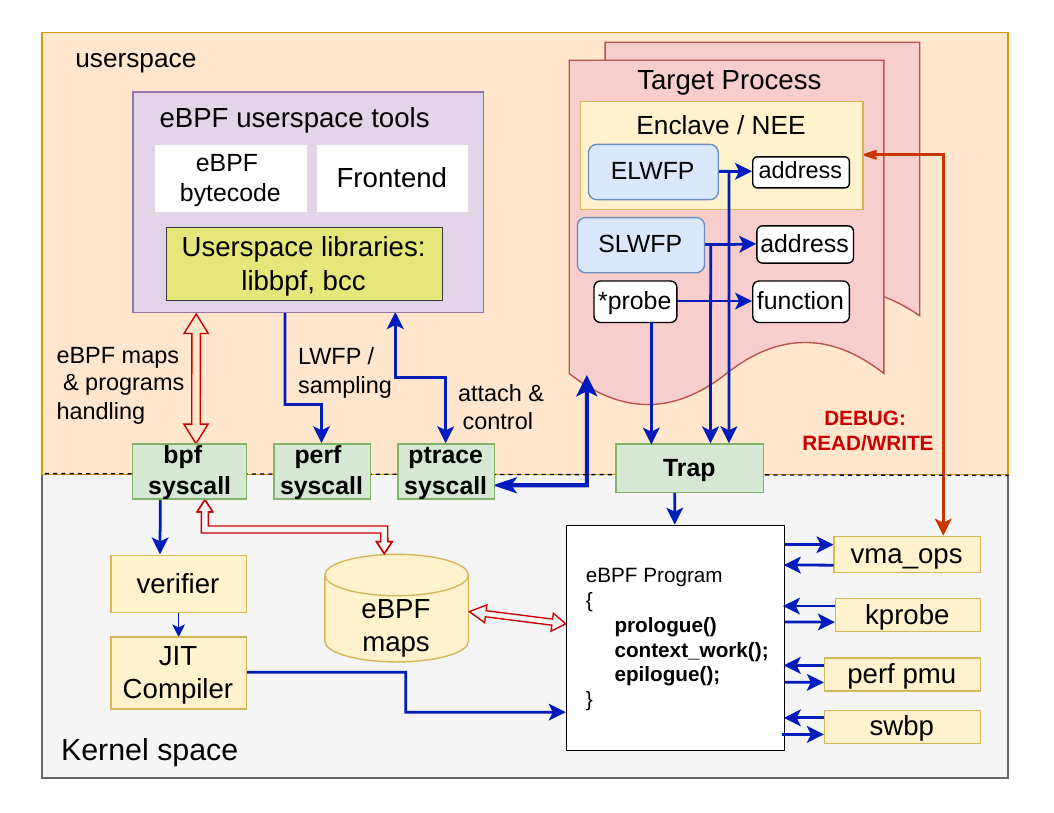}
    \caption{General overview of new tooling and interaction with tracees, eBPF, NEE and the new extensions.}
    \label{fig:general_structure}
\end{figure}

We present the design of several tools that demonstrate the enhanced capabilities afforded by the new extensions.
\cref{fig:general_structure} shows the new system layout with the added extensions and interaction between applications, NEE, eBPF and kernel.
The \textit{prologue} and \textit{epilogue} functions in eBPF summarize the context setup and updates needed in NEE, which is not the case with normal Linux probes.

The tools presented are as follows; a stack sampler ~\cref{subsubsection:stack-sampling}, dynamic and static tracing in \cref{subsubsection:general-design-tracers} and \cref{subsubsection:usdt-tool}, a syscall tracer \cref{subsection:strace-tool} and function timing \cref{subsubsection:function-timing}.

\subsubsection{CPU Profiling}\label{subsubsection:stack-sampling}
Also referred to as timed sampling or stack tracing.
The tool provides a CPU usage profile for an application to diagnose application hotspots.
An application is periodically interrupted and its stack traces are recorded and aggregated to 
show the fraction of execution time of each subcomponent.
Modern Linux distributions moved from stack frame pointers to the DWARF\cite{noauthor_DWARF_2017}\cite{eager_introduction_nodate}
format, to allow greater compiler optimization and software introspection.
DWARF specifies a virtual machine and its instructions which, when executed to termination, reproduce the desired cpu state at a desired location.

Compilers generate and append two DWARF location descriptor sections to binaries; \texttt{.eh\_frame(EH)} and an optional \texttt{.eh\_frame\_hdr(EFH)}.
EFH provides a compact header for the EH, enabling fast binary search of Function Description Entries (FDEs) during stack unwinding.
FDEs contains DWARF instructions for a given function.
Shared unwinding rules are contained in common instruction entry (CIE). More details for DWARF format and how to unwind can be found in \cite{noauthor_DWARF_2017}\cite{bastian_reliable_2019}.

\begin{figure}[hbtp]
    \centering
    \includegraphics[width=\linewidth]{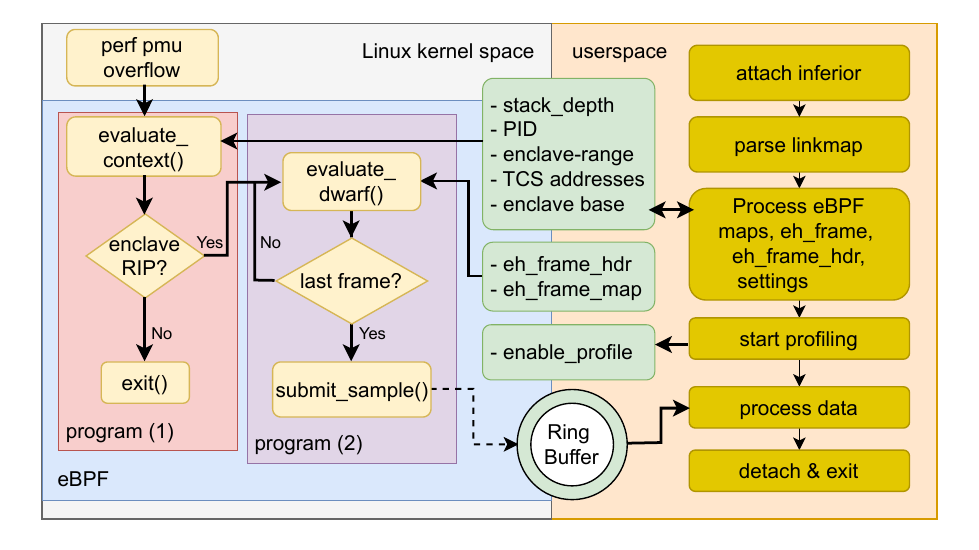}
    \caption{Overview of stack sampling application.}
    \label{fig:stack-sampling-overview}
\end{figure}

The design of the profiler is summarized in \cref{fig:stack-sampling-overview}.
It has a userspace frontend and an eBPF DWARF unwinder backend.
\noindent\textbf{Frontend} controls inferiors as described in \ref{subsubsection:controlling-target}.
Command line parameters include PID or inferior command, warm up time, sampling duration and frequency (100Hz default),
a linkmap containing library addresses and names loaded by inferior, and output files.
Parameters and inferior information are shared with eBPF code in maps, including EH and EFH number and data.
The eBPF-profiler program is loaded and associated with an \textit{fd} for a PEO periodic hardware timer as trigger. 
A ring buffer callback handler processes stack samples to resolve symbol names, aggregate and write the data to file.
Flame graphs are generated with scripts from \cite{gregg_flame_nodate}.

\noindent The \textbf{eBPF DWARF unwinder} consists of two programs.
\noindent(1) a \textbf{context evaluator} interprets the interruption context, matches configured values, and obtains
enclave context.
\noindent(2) \textbf{Unwinder} that does the DWARF VM emulation. It gets the enclave context, and then decodes and executes all CIE and FDEs, from the EH and EFH maps, to obtain and submit stack samples. 
It terminates on getting configured sample count, fully unwinding stack and on unwind errors. 

\subsubsection{General design of tracers}\label{subsubsection:general-design-tracers}
Tracers can be achieved with dynamic or static instrumentation.
Static instrumentation defines probe interfaces programmatically and some can be interpreted by the compiler to generate note sections describing them or symbol names following a certain naming convention. Two such cases are USDTs and exporting addresses of trace points with \texttt{nop} instructions.

Dynamic instrumentation can be applied to almost any instruction. However, the instructions need SSOL through emulation as described in \cref{subsubsection:single-stepping}.
Some common tasks for the tracers include.
(1) A mechanism to obtain the address of target probe and instrument it.
(2) Association of target probe to an eBPF program.
(3) eBPF program execution.
(4) Update inferior CPU state on eBPF program execution.

Handling (1) depends on target application e.g. Gramine-SGX implements ASLR while Intel SGX SDK does not.
As such, handling Gramine-SGX requires hooking uprobes into debug functions that report library loads to debuggers. In the other case, this value is knowable beforehand from the enclave binary.
(2) follows the description given in \cref{subsubsection:extending-perf-event-syscall}. (3) is dependent on the particulars of tracing task.
SSOL and context updates achieves (4). 

\begin{figure}[hbtp]
    \centering
    \includegraphics[width=\linewidth]{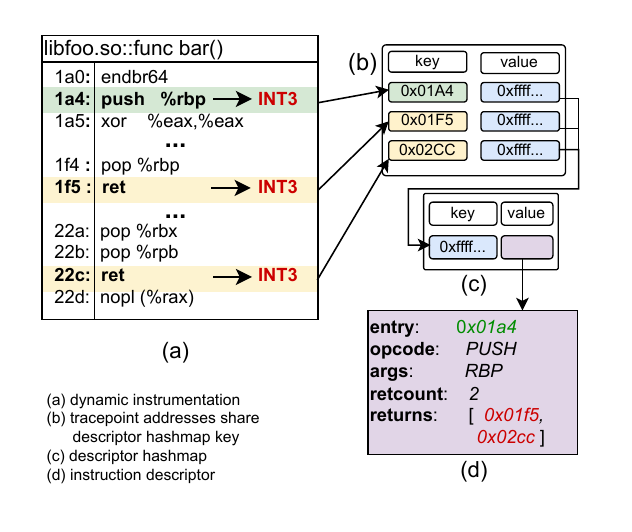}
    \caption{(a) Replacement of target instructions in function bar(). (b) Each replaced address shares the key to a map (c) whose value(d) is a struct describing the replaced instructions.}
    \label{fig:dynamic_instrumentation}
\end{figure}

Dynamic instrumentation of a function needs to group the function entry with all its return addresses.
This could be done as show in \cref{fig:dynamic_instrumentation}.
The tracepoint addresses, entry and all return addresses, of a function share a single instruction descriptor struct (d) stored in descriptor hashmap (c). 
The key in (c) is associated with all tracepoint addresses sharing the same instruction descriptor in hashmap (b). 
This way, the RIP at exception can be used to find the corresponding instruction descriptor.
A case statement can be used to call the correct emulation function on probe triggers.

A static tracer for USDTs is presented in \cref{subsubsection:usdt-tool}.
The timing tool \cref{subsubsection:function-timing} implements both static and dynamic tracing methods.
A system call tracer (strace) \cref{subsection:strace-tool} presents both a dynamic syscall tracer and profiler.

\subsubsection{Strace}\label{subsection:strace-tool}
This tool is designed to do  Linux syscall tracing following the System V supervisor call ABI for target LibOS.
The design follows structure in  \cref{subsubsection:context-management}-\cref{subsubsection:frontends} and  general ideas in \cref{subsubsection:general-design-tracers}.
The distinguishing design aspects are presented hereafter.

\textbf{Userspace frontend} has three unique characteristics. (1) Generation
of data structures describing arguments and sizes of each syscall in a shared map for
eBPF program to parse, collect and report syscall traces.
(2) It requires the information of all syscall entry functions in the target LibOS to hook into, which are usually LibOS specific.
(3) As dynamic tracer, it must define, determine and describe the specific binary instructions at all entry and exit points of syscall functions in order to prepare appropriate emulation functions. This is achieved through disassembly and inspection of the target instructions and appropriate locations to install the probes.
The loader instruments all target locations in the tracee process using ptrace or uprobes.
The Gramine-SGX support is done using uprobes and HB to dynamically detect enclave load addresses, as a way of mitigating variability from ASLR feature.
ESB are used to instrument the syscall function(s).

\textbf{eBPF backend} collects syscall traces from data structures stored in  syscall-indexed hashmap.
The main BPF function interprets syscall descriptions: argument counts, argument lengths, and the locations—registers or memory—from which to read them.
Traces are written to a shared ringbuffer.
Emulation of entry and exit instructions, and the resumption of enclave execution, follows \cref{subsubsection:single-stepping}.
To differentiate syscalls across threads, the userspace RBP serves as a LibOS thread identifier and as the key for temporary maps that store entry data.
When the syscall returns, the return value is appended to complete the trace.

\textbf{Data processing}. Syscall traces generated are chronologically ordered and stored to
file. The data is also aggregated to generate syscall profiles.

\subsubsection{Timing}\label{subsubsection:function-timing}
This tool uses dynamic instrumentation for benchmarking arbitrary functions and static instrumentation for probe benchmarking.
A struct defines each sample with the key identifier being the function address.
Like the previous tools, it needs to determine the addresses for each function and
the number of return functions as described in \cref{subsubsection:general-design-tracers}.
For function probing, it emulates only \texttt{push} and \texttt{ret} instructions with dynamic probes,  avoiding hard to emulate instructions like \texttt{endbr64}. This condition is ensured by the frontend.

\subsubsection{USDT}\label{subsubsection:usdt-tool}
This tool follows the same design as previous tools as far as frontend information for inferior
management is concerned. Data post processing is done in userspace.
Its unique characteristic is USDT parsing.
The userspace loader gets USDT information over command line or parsing the target library \textit{.note.stapsdt} section.

The target USDTs instructions are then executed and converted into structs that can be easily executed in the eBPF program.
This avoids implementing the full interpreter in eBPF.
SSOL for USDT is just incrementing the \texttt{RIP} and updating the context.
Specifically, this USDT tool is designed to handle MariaDB USDTs.
\section{Implementation}\label{section:implementation}
This section briefly details the implementation of features and tools that were presented in the \autoref{section:design}.

\textbf{Linux Kernel Extensions.}
The Linux kernel extensions were developed against kernel version 6.0.0, the earliest release providing the required eBPF capabilities, with 679 LoC.

\textbf{Dwarf Profiler}.
The profiler is implemented in C using libbpf\cite{noauthor_libbpflibbpf_2026} and code skeleton from  libbpf-bootstrap\cite{noauthor_libbpfbootstrap_nodate}.
Implementing an eBPF DWARF profiler is challenging due to verifier limits: programs have a ceiling of one million instructions, large EFH-map offsets are limited by eBPF addressing, loops explode instruction count, stack space is restricted to 512\,B, and call depth is limited to 31 frames. These constraints are mitigated using \textit{bpf\_loop} to handle large loops, \textit{bpf\_tail\_call} to partition large code, and map-backed temporary structs to minimize stack usage and store global context.
It has 2096 LoC of bpf, and 1277 LoC of userspace code.

\textbf{Strace Tool, Timing} and \textbf{USDT.}
These tools were implemented using BCC tools\cite{noauthor_iovisorbcc_2025}.
The bpf code is in C with LoC of 721, 672 and 560, respectively. The userspace frontend is written in Python 3.

General observations from implementation of the tooling include; shared userspace functionality, prologue and epilogue code in eBPF, 
and only minor distinctions for individual tools.
Userspace frameworks like \texttt{BCC} or tools like \textit{bpftrace}~\cite{noauthor_pftrace_nodate} normally automate the commonality in libraries or code generation, eliminating complexity for the developer.
This approach can be adopted to address the redundancy. 
\section{Evaluation and Applications}\label{section:evaluation-and-applications}
The tools described in \cref{subsection:tooling} were employed to evaluate applications executed on multiple SGX software stacks,
including the Intel SGX SDK\cite{noauthor_intelconfidential-computingsgx_nodate}, Gramine‑SGX\cite{noauthor_gramineprojectgramine_nodate}, and SCONE\cite{al_scone_2016}.
The timing tool measured the probe‑induced overheads. Collectively, these tools demonstrate the tracing, profiling, timing, and benchmarking dimensions of observability, with scenario‑specific details presented in each case.
\subsection{Experimental Setup}
All experiments are conducted on an Intel Xeon 6787P server with  16GB of EPC memory, 256GB RAM, and running  Ubuntu 24.04.2.
The benchmarks, tooling and applications run on a modified Linux 6.0.0 kernel for all measurements.
Reference benchmarks were made on a custom compiled stock Linux kernel 6.0.0.

\begin{figure*}[b!]
    \includegraphics[width=\textwidth]{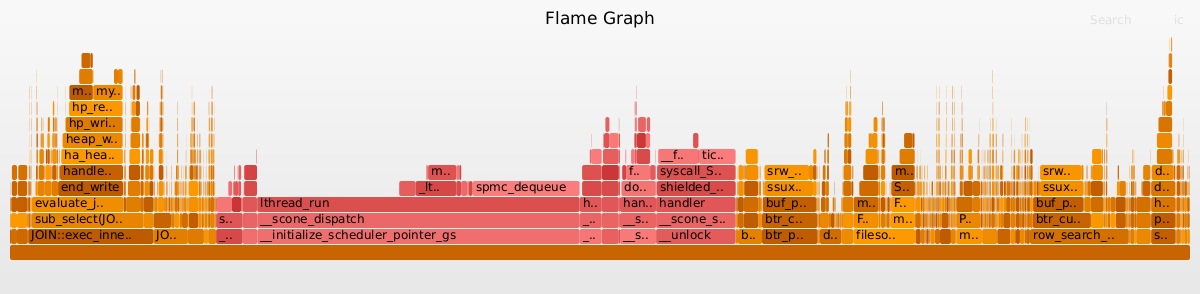}
    \caption{A flame graph of SCONE running MariaDB. The stack profiles appear in pink for SCONE and orange for MariaDB.}
    \label{fig:profiling}
\end{figure*}

\subsection{Probe Overheads}
\begin{figure}[hbtp]
    \centering
    \includegraphics[width=0.8\linewidth]{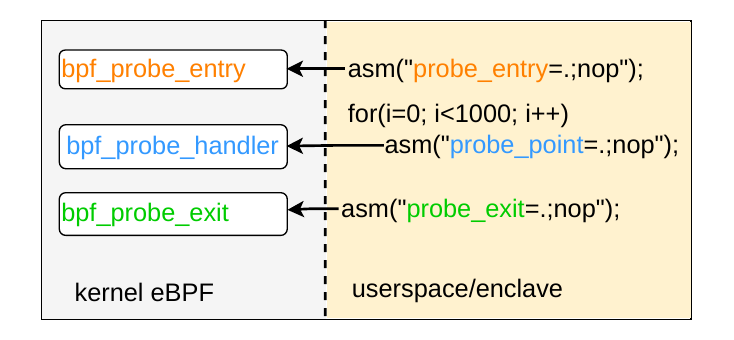}
    \caption{Probe overhead test plan and probe placement.}
    \label{fig:test-plan}
\end{figure}

This experiment measured the latency introduced by each probe.
ESB tests use Intel SGX SDK\cite{noauthor_intelconfidential-computingsgx_nodate} v2.27.
The test code was statically instrumented as shown in \cref{fig:test-plan}. HB probes are placed at the exit and entry for all tests.
The \textit{probe\_point} probe placement is as follows; ESB and HB for enclaves, SSB, HB and  uprobes for native modified kernel and only uprobes and HB with the stock kernel case.
CPU warm up code is run before invoking the experiment.
Timestamps are taken at the entry of \textit{bpf\_probe\_exit} and exit of \textit{bpf\_probe\_entry} probes. 
eBPF program at \textit{probe\_point} handles SSOL.
Further experiments were carried out to understand the latency of different components.

\cref{tab:probe-latency} shows the evaluated values for each probe type.
We observe that uprobes and HB have a 25\% and a 28\% higher latencies, respectively, over SSB. 
This is primarily due to the extra work uprobes kernel handlers do and the slow \#DB path for HB involving debug register management. SSB paths do not worsen the performance of uprobes relative to the stock kernel.
With a latency of 394 ns, SSB are a persuasive lightweight alternative to uprobes and HBs for typical native use cases as well as native NEE scenarios.

\begin{table}[hp]
\centering
\arrayrulewidth=0.8pt
\begin{tabular}{|l|c|c|c|}
\hline
\rowcolor{gray}
{\color{white} Probe} &
{\color{white} Stock} &
{\color{white} Native} &
{\color{white} Enclave} \\
\hline
uprobe & 509 & 505 & - \\
\hline
hw     & 477 & 492 & 2802 \\
\hline
sw     & -   & 394 & 2757 \\
\hline
\end{tabular}
\caption{Minimum probe latencies(ns) for new probes in enclaves and native code on modified Linux kernel, and stock kernel uprobes and hw probes. ESB is the sw probe in the Enclave. SSB is the sw in Native.}
\label{tab:probe-latency}
\end{table}

Both enclave probes, ESB and HB, show a similar minimum latency of $\sim2.8$ $\mu$s for an enclave$\rightarrow$eBPF$\rightarrow$enclave  round trip.
We measured an EENTER$\rightarrow$EXIT round trip as 1617 ns (6144 cycles at 3.8 GHz), indicating the enclave probes' latency is dominated by the AEX time.
Higher latencies should be expected and are proportional to the amount of processing done in the eBPF handler program.
Thus, designers should take the latencies into account.

\begin{figure*}[ht]
    \centering
    \includegraphics[width=\textwidth]{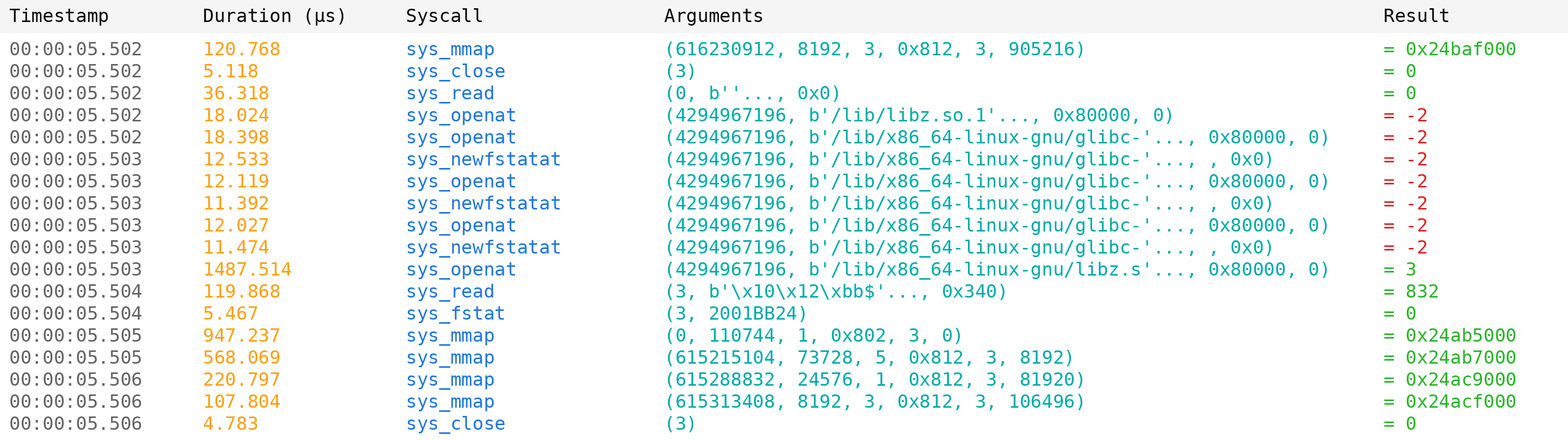}
    \caption{A brief strace snippet of Python3 running on Gramine‑SGX.}
    \label{fig:strace-log}
\end{figure*}

\subsection{Profiling}
Profiles give a clear, first‑hand view of an application’s performance and how computation is distributed across its internal functions or components.
The eBPF DWARF profiler described in \cref{subsubsection:stack-sampling} and \cref{section:implementation}  was used to profile MariaDB running on SCONE, at 100Hz, on the same host machine.
The client executed the Sysbench OLTP\cite{sysbench} read‑write database benchmarks on a separate machine and connected to the database remotely over the network.
GDB can also be used to get an enclave linkmap from a SCONE debug image.
The Flame graph in \cref{fig:profiling} was generated from the collected stack samples with a perl script from \cite{gregg_flame_nodate}. 

From \cref{fig:profiling}, enclave execution spends a significant amount of time in the SCONE scheduler (\(\_\_scone\_dispatch\)). In particular, function \textit{spmc\_dequeue} seems to take a considerable amount of execution time checking for new queue data\cite{arnautov_ffq_2017}.
This might cause a characteristic high CPU usage for the associated asynchronous syscall interface.
Since workload characteristics vary across applications, different runtime settings could be used to adapt the scheduler i.e it could have an IO heavy or compute heavy personality.
This profiler is general enough that, by providing relevant enclave information as described in \cref{section:implementation}, any LibOS can be profiled.
This generality is a testament to the utility of this work for tool creation. 
Prior to this work, profilers were either tied to a specific LibOS or embedded within a LibOS through dedicated introspection libraries.

\subsection{Tracing (eBPF Strace)}\label{subsection:application-tracing}
Tracing a syscall interface can provide a chronological order of executed system calls and the following trace information: syscall names, arguments, return values  and duration. 
The strace tool described in \cref{subsection:strace-tool} was used to hook and trace applications running on stock Gramine-SGX v1.9. 

From Python3 log shown in \cref{fig:strace-log}, it can be seen that several attempts to search and open \textit{libz} fail until the correct directory, \textit{\/lib/x86\_64-linux-gnu\/} .
Each failed attempt ends up invoking a \textit{sys\_newfstatat}, incurring unnecessary overheads.
Such failures are normally a result of incorrect environment variables or search path  misconfiguration.
A common performance optimization strategy is to reorder or customize the environment based on the installation path, prioritizing the locations that are most frequently accessed.

\begin{figure}[hbtp]
    \centering
    \includegraphics[width=\linewidth]{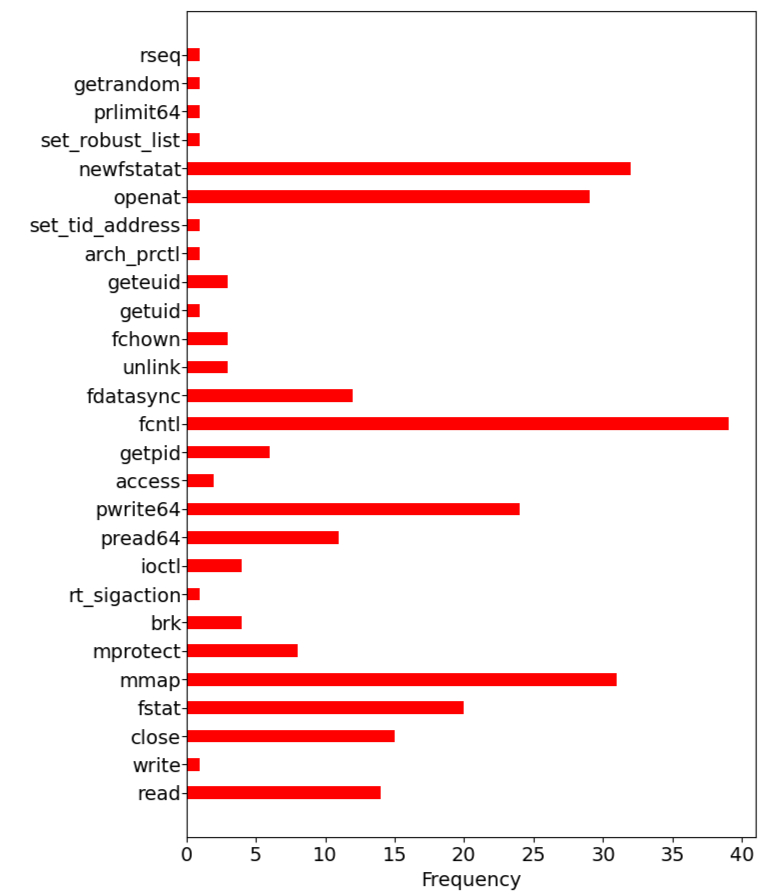}
    \caption{A syscall profile SQlite3 running under Gramine-SGX and executing provided sample sql code\cite{gramine_sqlite_example_v1_9}.}
    \label{fig:sqlite3-histogram}
\end{figure}

The syscall profile data for SQlite3 was rendered to generate a profile as shown in \cref{fig:sqlite3-histogram}.
A LibOS personality can be tuned depending on the syscall profile of target application to maximize performance.
Just like the failures observed in \cref{fig:strace-log}, large frequency of system calls can also be indicative of high frequency errors motivating further analysis or investigation.
This tool is also LibOS agnostic like the previous one.

\subsection{Function Timing }\label{subsection-probing}
\begin{figure}[h!]
    \includegraphics[width=0.85\linewidth]{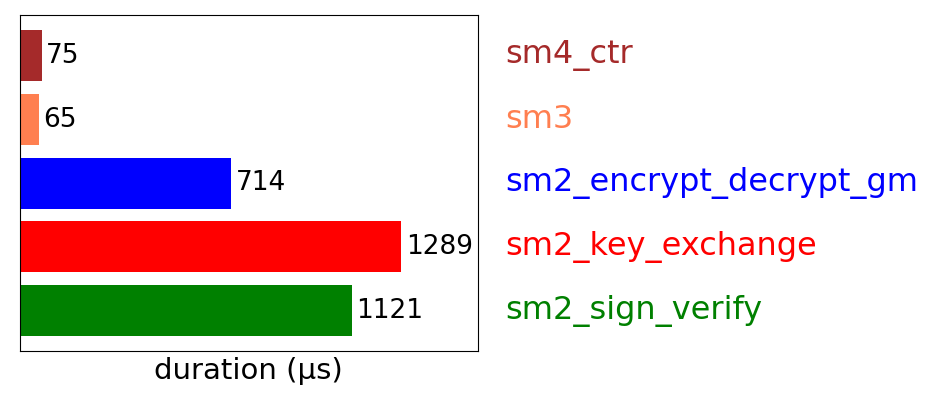}
    \caption{Latency of GM cryptography functions in Intel SGX SDK Sample Code\cite{intel_sgx_gmipp_sample_2025}.}
    \label{fig:functiontiming}
\end{figure}

The tool described in \cref{subsubsection:function-timing} was used to benchmark Intel SDK functions as shown in \cref{fig:functiontiming}, using dynamic instrumentation of the target functions.
The tool requires several command‑line parameters to measure enclave function latency: enclave base address, a list of function names, the target binary’s name and load address, and the addresses of the enclave TCS and SECS structures.

The timing tool can launch or attach to application to measure function latency.
This tool allows a developer or performance analyst to benchmark functions to understand  effects of code changes.

\subsection{USDTs}
\begin{figure}[hbtp]
    \includegraphics[width=0.34\textwidth]{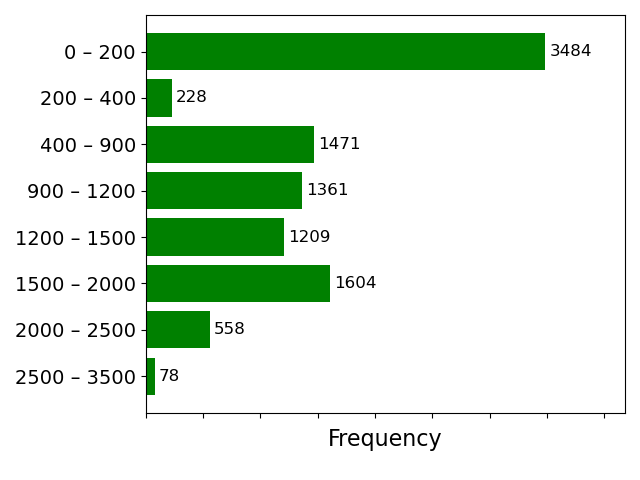}
    \caption{A histogram of MariaDB query latency ($\mu$s) with synthetic workoad from Sysbech OLTP read-write benchmark\cite{sysbench}.}
    \label{fig:mariadb-usdt}
\end{figure}

USDTs are employed in applications to provide stable tracing and benchmarking points unlike dynamic tracing.
This tool is inspired by \textit{dbslower}\cite{bcc_dbslower_3f5e402} tool from BCC.
MariaDB run under Gramine-SGX v1.19, with the setup and docker image adapted from \textit{enclaive} \cite{noauthor_enclaive-docker-mariadb-sgx_nodate} repository.
The tool instruments the following MariaDB v10.6 USDTs: \textit{query\_\_exec\_\_start, query\_\_exec\_\_done, query\_\_cache\_\_hit, query\_\_cache\_\_miss}.
Data captured by the tool was used to generate a histogram of query latencies for MariaDB as shown in \cref{fig:mariadb-usdt}.
The histogram shows a high number of short queries, likely showing good caching performance.
This analysis can also be applied to complex setups to optimize database performance.

Similar BCC tools can be ported for NEE observability.
The tool includes a frontend for SCONE that manages enclave related metadata.
This is equivalently extendable to other LibOSes.

\section{Discussion}\label{section:discussion}
We have demonstrated the flexibility of our primitives with Intel SGX, LibOSes, and native cases.
New NEE architectures can be supported by implementing the driver interface and extending PEO to cater for specific characteristics. 
For instance, a TEE like vSGX\cite{zhao_vsgx_2022} implements SGX ISA and thus integration is straight forward. 
A more general approach to support any hardware architecture might be done with a PEO probe that can flexibly match any number of registers, but this is left for future work.
With the most limiting problems solved by this work, a simpler approach to this task would be utilization of \textit{kfuncs}~\cite{kfuncs} and \textit{kprobes}, along the new TEE's exception path, and use eBPF for context matching.

eBPF tools are characteristically shorter, faster and easier to write than equivalent \textit{ptrace}-based tools.
As such, existing tooling can be easily ported to support NEEs.
For instance, several diverse tools from BCC\cite{noauthor_iovisorbcc_2025} have the core logic in eBPF.
Introducing an NEE prologue, epilogue and a corresponding context structure would be sufficient for overwhelming majority of applications.
For userspace code, such tools can adopt a plugin approach to cater to specific NEEs.

\section{Related Work}\label{section:related-work}

\textbf{eBPF Extensions to the Linux Kernel.}
There are several efforts to extend eBPF interfaces. 
\textit{BPF struct\_ops}~\cite{structops} enables prototyping kernel extensions in eBPF and calling them from within the kernel. We do the opposite.
\textit{kfuncs}~\cite{kfuncs} allow eBPF programs to call kernel functions and approach similar capabilities but require driver-specific context awareness.

\textbf{Userspace eBPF Runtimes.}
Userspace runtimes relocate eBPF execution outside the kernel. 
Bpftime\cite{zheng_bpftime_2023} moves uprobes to userspace but depends on shared memory and \texttt{ptrace}, limiting usage hardware NEEs. 
\textit{uBPF}\cite{noauthor_iovisorubpf_2024} and \textit{rbpf}\cite{monnet_qmonnetrbpf_2024} provide embeddable eBPF VMs and support limited NEE scenarios, but do not address broader observability challenges.

\textbf{Specialized TEE Tooling.}
TEE-specific tools focus on narrow or LibOS-specific scenarios. 
TEEMon~\cite{krahn_teemon_2020} monitors SGX and Linux metrics without inspecting enclave execution. 
SGX-perf~\cite{weichbrodt_sgx-perf_2018}, TEE-perf~\cite{bailleu_tee-perf_2019}, and TS-Perf~\cite{suzaki_ts-perf_2021} rely on instrumentation or recompilation. 
SGX-Tuner~\cite{mazzeo_sgxtuner_2022} and SGX LibOSes ~\cite{noauthor_gramineprojectgramine_nodate,al_scone_2016} expose only limited internal metrics. 
SGXoMeter~\cite{mahhouk_sgxometer_2021} benchmarks Intel SDK enclaves, while GDB SGX extensions~\cite{noauthor_sgx_gdbplugin} and Intel VTune~\cite{intel_vtune_2026} support debugging or hotspot profiling. 
These tools are single-purpose, enclave-specific, or require recompilation, and cannot generically observe NEEs.

\textbf{Profiling.}
X-Prof~\cite{eunomia_xpu_perf} provides DWARF-based profiling for CPUs and GPUs but does not support SGX debug or production enclaves. 
DWARF unwinding optimizations~\cite{bastian_reliable_2019} improve stack unwinding performance. 
Native \texttt{perf}~\cite{noauthor_perftool_nodate} can measure profile native execution not NEE internals. 
The DWARF profiler in this work demonstrates the capabilities of our primitives rather than optimized profiling performance.

\textbf{Dynamic Instrumentation.}
Native dynamic instrumentation tools such as DTrace~\cite{cooper_dtrace_2012}, Utrace~\cite{keniston_ptrace_nodate}, and kprobes~\cite{mavinakayanahalli_probing_nodate} are effective for standard Linux applications but cannot operate inside NEEs due to restricted interfaces and non-standard execution paths. 
Ratel~\cite{ratel} adapts DynamoRIO to SGX for full instruction-level interposition. 
The primitives introduced in this work generalize such capabilities across diverse NEEs.

\textbf{Hardware Tracing Comparison.}
Hardware tracing mechanisms such as Intel PT and LBR provide exhaustive control‑flow visibility but generate large traces, whereas a software‑driven primitive enables selective, condition‑based state capture suitable for NEEs.

\section{Conclusion}\label{section:conclusion}
In this work presented a set of primitives that solve observability problem of NEEs from eBPF with SGX serving as the quintessential example.
Linux kernel extensions to virtual memory areas and the driver file‑operations interface were developed to allow safe access to NEE memory.
Together, these primitives enable the development of rich tooling such as profilers and tracers. They provide the necessary flexibility and generality to support a wide range of observability use cases. The flexibility and generality of this work forms a solid base upon which diverse and capable tooling can be developed to close the observability gap with native tooling.

\bibliographystyle{ACM-Reference-Format}
\bibliography{sgxdebugebpf}

\end{document}